\documentclass[aps,prd,twocolumn,superscriptaddress,showpacs]{revtex4}
\usepackage{graphicx}
\usepackage{amsmath}
\usepackage{amsfonts}
\usepackage{amssymb}
\usepackage{amssymb}
\usepackage{slashed}
\usepackage{pstricks,pst-coil}
\usepackage{dsfont}
\usepackage{bm}

\begin{document}
\title{Electric birefringence in Euler-Heisenberg pseudo-electrodynamics}
%
%
%\author{Rodrigo Turcati}\email{rturcati@sissa.it}

%\affiliation{SISSA, Via Bonomea 265, 34136 Trieste, Italy and \\
%                 INFN, Sezione di Trieste, Via Valerio 2, 34127 Trieste, Italy}
%\affiliation{Departamento de F\'isica e Qu\'imica, Universidade Federal do Esp\'irito Santo, Av. Fernando Ferrari, 514, Goiabeiras, 29060-900, Vit\'oria, ES, Brazil}
%\affiliation{Laborat\'orio de F\'isica Experimental (LAFEX), Centro Brasileiro de Pesquisas F\'isicas (CBPF), Rua Dr Xavier Sigaud 150, Urca, 22290-180 Rio de Janeiro, Brazil}

\author{M. J. Neves}\email{mariojr@ufrrj.br}

\affiliation{Departamento de F\'isica, Universidade Federal Rural do Rio de Janeiro, BR 465-07, 23890-971, Serop\'edica, Rio de Janeiro, Brasil.}
\begin{abstract}
The fermion sector of the pseudo-quantum electrodynamics is integrated functionally to generate a non-linear electrodynamics, that it is called 
Euler-Heisenberg pseudo-electrodynamics. A non-local Chern-Simons topological term is added to the original 
lagrangian of the pseudo-quantum electrodynamics in which a most complete electrodynamics gauge invariant in 1+2 dimensions is proposed.
As consequence of the fermionic sector, we obtain a non-linear contribution in the electromagnetic fields 
that breaks the Lorentz symmetry due to Fermi velocity. From the Euler-Heisenberg pseudo-electrodynamics, we study the properties of the plane 
wave propagating in a planar medium under an uniform and constant electromagnetic background field. 
The properties of the planar material are discussed through the electric permittivity tensor and magnetic permeability, that are functions of the frequency, wavelength and of the background fields. The dispersion relations and the refractive index are calculated in the presence of a uniform magnetic field, and also in the case only of an electric background field. The birefringence phenomenon emerges only when the electric background field is considered.

\end{abstract}
%\pacs{14.70.-e, 12.60.Cn, 13.40.Gp}
\maketitle

% \begin{keyword}
% higher-derivative electrodynamics; complex-shell renormalization; anomalous magnetic moment of the electron; Uehling potential; renormalization group
% %% keywords here, in the form: keyword \sep keyword
%
% %% MSC codes here, in the form: \MSC code \sep code
% %% or \MSC[2008] code \sep code (2000 is the default)
%
% \end{keyword}

%\end{frontmatter}

% \linenumbers

\section{Introduction}
%\label{}
%

%
Field theories in $1+2$ dimensions have attracted great interest due to it application in condensed matter physics.
Materials like graphene in connection with the quantum Hall effect \cite{Gorbar,Gusynin,Herbut,Herbut2,Laughlin,Chamon,Ando}, 
topological isolators \cite{Qi,Hasan,Chiu,Zhao,Qi2008}, superconductivity in layered materials \cite{Tesanovic,Zhang,Franz,Kivelson,Marino2018} 
have showed that quantum field theories (QFTs) provide an excellent theoretical description for experimental results in material physics. In 
Dirac materials, electrons move at Fermi velocity through a semi-relativistic dynamics are good material class to test QFTs \cite{Castro}. 
The interactions between electrons in planar materials are mediated by a gauge field theory known as pseudo-electrodynamics (PED) \cite{Marino93}. 
The PED is a non-local Abelian gauge theory obtained confining the classical sources of the usual Maxwell electrodynamics to a spatial plane, 
that results in the dimensional reduction to $1+2$ dimensions. When the PED is coupled to fermions through a $U(1)$ local gauge invariance, 
the model is called pseudo-quantum electrodynamics (PQED). Important properties of the PQED were demonstrated as the 
canonical quantization, causality, unitarity and renormalization \cite{Amaral,Marino2014}. These properties are fundamentals for the PQED as a possible QFT 
in the description of quantum phenomenon in planar materials. Several extensions of the PQED also were studied through the QFT 
approach, as the addition of Chern-Simons topological term \cite{Alves,Alves2}, the Proca pseudo-electrodynamics \cite{Ozela,Ozela2,VanSergio,Xing}, 
the Lee-Wick pseudo-electrodynamics \cite{MarioPRD2025,MJNevesPLB2025,MJNevesEPJC2026}, and the supersymmetry \cite{Petrov}. 
In all these references, different approaches are discussed to include mass to the pseudo-gauge field. 
In parallel to the success of the PQED, investigations of non-linear effects from the 
quantum electrodynamics applied to Dirac material emerges from polarized vacuum Dirac sea, see the refs. \cite{Keser,Jorge}. 
These non-linear effects are characteristics of the very strong fields found in neutrons stars or in heavy ion colliders. 
Historically, non-linear electrodynamics (NLEDs) emerged for the first time with M. Born and L. Infeld to explain the static electric field 
of a point charge at origin \cite{Born}. Some years after, H. Euler and W. Heisenberg obtained the fermionic effective action 
under an external EM field at one loop in the quantum electrodynamics (QED) \cite{EulerHeisenberg}. In present days, others NLEDs has been 
proposed with several applications in many research areas as Cosmology, alternative Gravitation theories, material physics and others, see the refs.  \cite{Sorokin1,Sorokin2,Sorokin3,GENBI,LOG,Gaete_EPJC_2022,MJNevesPRD2021,MJNevesPRD2023,MJNevesJPA2023}. The observation of non-linear 
effects also appear in the Hall effect \cite{Qiong,Kang,Fu}, and in topological photonics \cite{Daria}. Effects of the QED has been studied 
in the graphene, as the Schwinger effect \cite{Allor,Dora,Mostepanenko,MacLean,Akal,Golub,Francois}. In Dirac materials, the non-linear electromagnetism 
is studied as response to transport properties with applications to photovoltaic \cite{Culcer,Sodeman,SodemanPRL,Rostami1,Rostami2,Bhalla}. In the graphene, the dispersion of electrons leads to strong nonlinear electromagnetic response of material that is described by a third order NLED \cite{Mikhailov}. The results of all these references 
motivate us to explore the non-linear ED that arises from the radiative corrections at one loop in the PQED. This non-linear and planar 
ED has not been explored in the literature yet.         
In this paper, we calculate the effective action for the fermions in the PQED coupled to the pseudo-electromagnetic 
field in the presence of a non-local topological CS term. The result is the so called non-linear Euler-Heisenberg pseudo-electrodynamics (EHPED). This non-linear 
electrodynamics has a Lorentz symmetry breaking coming from the fermionic sector that introduces naturally the Fermi velocity in the spatial components of the Dirac matrices.
We study the consequences of the EHPED when the model is submitted to a uniform and constant electromagnetic background field in $1+2$ dimensions. This is an important point to explore the characteristics of the material medium governed by the EHPED. Thereby, 
we investigate the plane wave solutions propagating in a planar medium ruled by the EM background field. The optical properties of the planar material medium are discussed through the electric permittivity tensor, and of the magnetic permeability, that is a function scalar in $1+2$ dimensions. We obtain the wave equation in the frequency space whose solutions yield the dispersion relations and the refractive index as functions of the electron mass, Fermi velocity, wave frequency, and of the background EM fields. The solutions for the refractive index show the conditions for a dispersive medium, as the dispersion and absorption of EM waves. As application of these results, the optical phenomenon of the birefringence is discussed in terms of the EM background field for the EHPED.
The paper is organized as follows :  In the section (\ref{sec2}), the non-linear Euler-Heisenberg pseudo-electrodynamics is presented. In the section (\ref{sec3}), we introduce the linearization of the model using the prescription of an external electromagnetic field. In this same section, we obtain the refractive index for plane wave solutions. The section (\ref{sec4}) is dedicated to discussion of the birefringence. For end, in the section (\ref{sec5}), we highlight the conclusions.  
We consider the natural units with $\hbar=c=1$ throughout paper.  The metric signature is $\eta_{\bar{\mu}\bar{\nu}}=\mbox{diag}(+1,-1,-1)$, 
in which the bar index $\bar{\mu}=\{0,1,2\}$ are adopted for vectors and tensors in $1+2$ dimensions.
The electric and magnetic fields have the following conversions of $\mbox{eV}^2$ to V/cm and Tesla (T) in natural units : 
$1 \, \mbox{eV}^2=1.54 \times 10^{4} \, \mbox{V/cm}$ and $1 \, \mbox{eV}^2 = 692.5 \, \mbox{T}$, respectively. 
%

%
%**************************************************************************************************

\section{The Euler-Heisenberg pseudo-electrodynamics}
\label{sec2}
Massive fermions in $1+2$ dimensions are coupled to the pseudo-electrodynamics by the lagrangian density \cite{Fuhrer,Riazudin,Keser,Alves2} :
\begin{eqnarray}\label{PQED}
\mathcal{L}_{PQED} &=& \bar{\psi}\left(i\,\Gamma^{\bar{\mu}}D_{\bar{\mu}}-\Delta\,\mathds{1}\right)\psi
%-e\,\bar{\psi}\,\Gamma^{\bar{\mu}}A_{\bar{\mu}}\,\psi
-\frac{1}{4}\,F_{\bar{\mu}\bar{\nu}} \, \frac{2}{\sqrt{\bar{\Box}}} \,F^{\bar{\mu}\bar{\nu}}
\nonumber \\
&&
+\frac{\theta}{2} \, \epsilon^{\bar\mu\bar\nu\bar\rho} A_{\bar{\mu}} \frac{2}{\sqrt{\bar{\Box}}} \, \partial_{\bar{\nu} } A_{\bar\rho} \; ,
\end{eqnarray}
where $\psi$ is a Dirac spinor of two components, $\Delta=0.1$ eV is the gap over two of the planar material. For many examples of Dirac isolators and graphene, 
we have the gap of $2\Delta \simeq 2$ eV.    
%with $m=0.5$ MeV being the electron mass, 
$\Gamma^{\bar{\mu}}=(\gamma^{0},v_F\gamma^{i})\,(i=1,2)$ are the Dirac matrices in which the Fermi velocity $(v_F)$ emerges in the spatial component of the $\gamma^{\bar{\mu}}$-matrices, 
that satisfy the relation $\gamma^{\bar{\mu}}\gamma^{\bar{\nu}}=\eta^{\bar{\mu}\bar{\nu}}+i \, \sigma^{\bar{\mu}\bar{\nu}}$, and 
$\bar{\psi}=\psi^{\dagger}\,\gamma^{0}$ is the adjunct field. The covariant derivative operator is $D_{\bar{\mu}}=\partial_{\bar{\mu}}+ie\,A_{\bar{\mu}}$, 
in which the fundamental charge $(e)$ is related to fine structure constant $(\alpha=1/137)$ by $e^2=4\pi \, \alpha\,v_F$ in natural units. The notation for 
derivative operator is such that $\partial_{\bar{\mu}}=(\partial_{t},\partial_{x},\partial_{y})$, and the D'Alambertian operator in $1+2$ dimensions is $\bar{\Box}=\partial_{\bar{\mu}}\partial^{\bar{\mu}}=\partial_{t}^2-\partial_{x}^2-\partial_{y}^2$. The matrices $\gamma^{\bar{\mu}}=(\gamma^{0},\gamma^{1},\gamma^{2})$ 
are represented in terms of the Pauli matrices $(\sigma_{1},\sigma_{2},\sigma_{3})$ as
\begin{eqnarray}
\gamma^{0}&=&\sigma_{3}=
\left[
\begin{array}{cc}
1 & 0 \\
0 & -1 \\
\end{array}
\right]
\; , \;
\gamma^{1}=i\sigma_2=\left[
\begin{array}{cc}
0 & 1 \\
-1 & 0 \\
\end{array}
\right]
\; , \;
\nonumber \\
\gamma^{2}&=&-i\sigma_{1}=\left[
\begin{array}{cc}
0 & -i \\
-i & 0 \\
\end{array}
\right] \; .
\end{eqnarray}
%
%in which $\sigma_{i}=(\sigma_{1},\sigma_{2},\sigma_{3})$ are the Pauli matrices. 
In the gauge sector of (\ref{PQED}), $F_{\bar{\mu}\bar{\nu}}=\partial_{\bar{\mu}}A_{\bar{\nu}}-\partial_{\bar{\nu}}A_{\bar{\mu}}=(-E^{i},-\epsilon^{ij} B)$ 
is the EM field tensor, with the $3$-potential $A^{\bar{\mu}}=(V,A^{i})\,(i=1,2)$. We have added a non-local Chern-Simons (CS) term in which 
$\theta$ is a parameter with mass dimension. The action correspondent to the lagrangian (\ref{PQED}) is $U(1)$-gauge invariant 
as it can be checked by the gauge transformations on the $3$-potential, and also on the local phase transformation in the spinor field. 
In quantum field theory approach, the generating functional for fermions 
interacting minimally with the Chern-Simons-Maxwell pseudo-ED associated with the lagrangian (\ref{PQED}) is 
%
%\begin{widetext}
\begin{equation}\label{FunctionalZ}
Z[J^{\bar{\mu}}]=\int \mathcal{D}A^{\bar{\mu}} \, \mathcal{D}\bar{\psi} \, \mathcal{D}\psi \,
%\nonumber \\
\mbox{exp} \left[i \, S_{PQED} + i \! \int d^3x \,
J_{\bar{\mu}} A^{\bar{\mu}} 
%)  
\right] \; ,
\end{equation}
%\end{widetext}  
%
where $S_{PQED}$ is the functional action of (\ref{PQED}), and $J^{\bar{\mu}}$ is the classical source of the gauge field, respectively.
Here, we are considering the external source for fermions are nulls. Making the functional integrations on the fermion fields $\bar{\psi}$ 
and $\psi$, the generating functional is reduced to
\begin{eqnarray}\label{FunctionalZA}
Z[J^{\bar{\mu}}] &=& \int \mathcal{D}A^{\bar{\mu}} \,
%\nonumber \\
\mbox{exp} \left[ i \! \int d^3x \left( {\cal L}_{PED} +
J_{\bar{\mu}} A^{\bar{\mu}} \right) \,
\right] \times
\nonumber \\
&&
\times \, \det(i\,\Gamma^{\bar{\mu}}D_{\bar{\mu}}-\Delta\,\mathds{1}) \; ,
\end{eqnarray}
in which ${\cal L}_{PED}$ is the gauge sector of (\ref{PQED})
\begin{eqnarray}
{\cal L}_{PED}=-\frac{1}{4}\,F_{\bar{\mu}\bar{\nu}} \, \frac{2}{\sqrt{\bar{\Box}}} \,F^{\bar{\mu}\bar{\nu}}+\frac{\theta}{2} \, \epsilon^{\bar\mu\bar\nu\bar\rho} A_{\bar{\mu}} \frac{2}{\sqrt{\bar{\Box}}} \, \partial_{\bar{\nu} } A_{\bar\rho} \; .
\end{eqnarray}
Using the identity $\det({\cal O})=\exp\left\{\mbox{Tr}\left[\mbox{tr}\ln({\cal O})\right]\right\}$ for the operator ${\cal O}=i\,\Gamma^{\bar{\mu}}D_{\bar{\mu}}-\Delta\,\mathds{1}$,
in which $\mbox{tr}$ is the trace over the coordinate space in $1+2$ dimensions, and $\mbox{Tr}$ symbols the trace over the $\Gamma^{\bar{\mu}}$-matrices. Thereby, the generating functional (\ref{FunctionalZA}) is written as
\begin{equation}\label{FunctionalZA}
Z[J^{\bar{\mu}}] = \int \mathcal{D}A^{\bar{\mu}} \,
%\nonumber \\
\mbox{exp} \left[ \, i \int d^3x \left( {\cal L}_{PED} + {\cal L}_{eff} +
J_{\bar{\mu}} A^{\bar{\mu}} \right) \right] \; ,
\end{equation} 
where the effective lagrangian is defined by
\begin{eqnarray}\label{Leff}
{\cal L}_{eff}=-i\,\mbox{Tr}\left[ \, \langle x| \ln(i\,\Gamma^{\bar{\mu}}D_{\bar{\mu}}-\Delta\,\mathds{1}) | x \rangle \, \right] \; . 
\end{eqnarray}
We calculate the derivative of (\ref{Leff}) in relation to $\Delta^2$ to obtain the expression 
\begin{eqnarray}\label{LeffmTr}
\frac{\partial {\cal L}_{eff}}{\partial \Delta^2} &=& \frac{i}{2\Delta} \, \mbox{Tr}\left[ \, \langle x| \frac{1}{i\,\Gamma^{\bar{\mu}}D_{\bar{\mu}}-\Delta\,\mathds{1}} | x \rangle \, \right] 
\nonumber \\
&=& -\frac{i}{2\Delta}\,\mbox{Tr}\left[ \, \langle x| \frac{i\,\Gamma^{\bar{\mu}}D_{\bar{\mu}}+\Delta\,\mathds{1} }{(\Gamma^{\bar{\mu}}D_{\bar{\mu}})^2+\Delta^2} | x \rangle \, \right] 
\; .
\end{eqnarray}
Using that the trace of $\Gamma^{\bar{\mu}}$ is null, the expression (\ref{LeffmTr}) is simplified as
\begin{equation}\label{Leffm}
\frac{\partial {\cal L}_{eff}}{\partial \Delta^2} = -\frac{i}{2} \, \mbox{Tr}\left[ \langle x| \frac{ 1 }{(\Gamma^{\bar{\mu}}D_{\bar{\mu}})^2+\Delta^2} | x \rangle \right] \; ,
\end{equation}
in which the properties of $\Gamma^{\bar{\mu}}$ allow to write the denominator as 
\begin{eqnarray}
(\Gamma^{\bar{\mu}}D_{\bar{\mu}})^2=D_{\bar{\mu}}D^{\bar{\mu}}+iev_F({\bm \sigma}\cdot{\bf E}+i v_F B\sigma_3) \; .
\end{eqnarray}
The fraction of (\ref{Leffm}) is written in terms of the Schwinger's integral 
\begin{eqnarray}
\frac{1}{A-i\epsilon}=i\int_{0}^{\infty} ds \; e^{-is(A-i\epsilon)} \; ,
\end{eqnarray}
where $s$ is the proper-time parameter, and $\epsilon$ is a positive real parameter. Thus, we obtain
\begin{eqnarray}\label{Leffmints}
\frac{\partial {\cal L}_{eff}}{\partial \Delta^2} &=& \frac{1}{2} \int_{0}^{\infty} ds \, e^{-is(\Delta^2-i\epsilon)} \, \mbox{Tr}[ \, e^{sev_F({\bm \sigma}\cdot{\bf E}+iv_F B\sigma_3)}\,]
\nonumber \\
&&
\times \langle x | e^{-is(D_{\bar{\mu}}D^{\bar{\mu}})} |x\rangle  \; .
\end{eqnarray}
Integrating in relation to $\Delta^2$, the expression for the effective lagrangian is 
\begin{eqnarray}\label{LeffTr}
{\cal L}_{eff} &=& \frac{i}{2} \int_{0}^{\infty} \frac{ds}{s} \, e^{-is(\Delta^2-i\epsilon)} \, \mbox{Tr}[ \, e^{sev_F({\bm \sigma}\cdot{\bf E}+iv_F B\sigma_3)}\,]
\nonumber \\
&&
\times \langle x | e^{-is(D_{\bar{\mu}}D^{\bar{\mu}})} |x\rangle \; .
\end{eqnarray}
The trace over the exponential of the Pauli matrices is 
\begin{eqnarray}\label{Trresult}
\mbox{Tr}[ \, e^{sev_F({\bm \sigma}\cdot{\bf E}+i v_F B\sigma_3)}\,]=2\,\cosh(sev_F\sqrt{{\cal F}}) \; , \, 
%\mbox{if} \, |{\bf E}|>B
%\\
%\mbox{Tr}[ \, e^{se({\bm \sigma}\cdot{\bf E}+i B\sigma_3)}\,]=2\,\cosh(se\sqrt{{\cal F}}) \, , \, \mbox{if} \, |{\bf E}|>B
\end{eqnarray}
where we have defined ${\cal F}={\bf E}^2-v_F^2B^2$. The result (\ref{Trresult}) is valid if $|{\bf E}|>v_{F}B$. When $|{\bf E}|<v_{F}B$,
the hyperbolic cosine transforms in a cosine function. The term in the second line of (\ref{LeffTr}) in $1+2$ dimensions is given by 
\cite{McArthur}
\begin{equation}
\langle x | e^{-is(D_{\bar{\mu}}D^{\bar{\mu}})} |x\rangle=\frac{1}{8 \pi s \, \sqrt{i\pi s}}\,\sqrt{\det\left[ \frac{seG}{\sinh(seG)} \right]} \; ,  
\end{equation}
where $G$ sets the $3\times 3$ matrix :
%associated with the tensor $F^{\mu\nu}$:
%
\begin{eqnarray}\label{matrixG}
G=
\left(
\begin{array}{ccc}
0 & -E_{x} & -E_{y} \\
E_{x} & 0 & -v_F B \\
E_{y} & v_F B & 0 \\
\end{array}
\right) \; .
\end{eqnarray}
Notice that the $G$-matrix is analogous to antisymmetric matrix associated with the $F^{\bar{\mu}\bar{\nu}}$-tensor in which the magnetic field component 
is multiplied by the Fermi velocity. Substituting all these results in (\ref{LeffTr}), the effective lagrangian is
\begin{eqnarray}\label{Leffresult}
{\cal L}_{eff} &=& \frac{\sqrt{i}}{8\pi^{3/2}}\!\int_{0}^{\infty} \! \frac{ds}{s^{5/2}} \, e^{-is \Delta^2} \cosh\left(sev_F\sqrt{{\cal F}}\right)
\nonumber \\ 
&&
\times \, 
\sqrt{\det\left[ \frac{seG}{\sinh(seG)} \right]} \, .
\end{eqnarray}
In the weak field approximation, the relevant terms up to quartic order in the EM field are read as
\begin{eqnarray}
&&
\cosh(sev_F\sqrt{{\cal F}}) \,
\sqrt{\det\left[ \frac{seG}{\sinh(seG)} \right]} 
\simeq 1+\frac{e^2s^2}{12}(6v_F^2-1)\,{\cal F}
\nonumber \\
&&
+\frac{e^4s^4}{24}\left(v_F^4-v_F^2+\frac{3}{20}\right)\,{\cal F}^2
+{\cal O}({\cal F}^3) \; ,
\end{eqnarray}
that substituting in the integral (\ref{Leffresult}), the first term is divergent in $s=0$. Therefore, 
we need to introduce a renormalization scheme in the effective lagrangian (\ref{Leffresult}). The regulator 
parameter $(s_0)$ is introduced in the lower limit of the $s$-integral, and the counter-term needed for renormalization is given by
\begin{eqnarray}
&&
{\cal L}_{eff}^{(c)}(s_0)= \frac{-\sqrt{i}}{8\pi^{3/2}}\!\int_{s_0}^{\infty} \! \frac{ds}{s^{5/2}} \, e^{-is \Delta^2} 
\nonumber \\
&&
\simeq\frac{-1}{12\pi s_{0}(-i\pi s_0)^{1/2}}+\frac{i \, \Delta^2}{4\pi (-i\pi s_{0})^{1/2}}+\frac{\Delta^3}{6\pi}
%\!\left[1+\frac{e^2s^2}{12}(6v_F^2-1)\,{\cal F}\right] 
\, ,
\end{eqnarray}
for $s_{0} \ll s$. Thus, when $s_{0}\rightarrow 0$, the two first terms are divergent and the last term is non-physical that also will be removed 
in the renormalized lagrangian. The renormalized Euler-Heisenberg lagrangian density is written as
\begin{widetext}
\begin{eqnarray}\label{LEH}
{\cal L}_{EH}=\frac{\sqrt{i}}{8\pi^{3/2}}\lim_{s_0\rightarrow 0} \int_{s_0}^{\infty} \! \frac{ds}{s^{5/2}} \, e^{-is \Delta^2}
\left[ \, \cosh(sev_F\sqrt{{\cal F}}) \, \sqrt{\det\left[ \frac{seG}{\sinh(seG)} \right]}-1
%-\frac{e^2s^2}{12}(6v_F^2-1)\,{\cal F} 
\, \right] \; .
\end{eqnarray}
\end{widetext}
For weak EM fields, the physical lagrangian (\ref{LEH}) in terms of the fine structure constant is
\begin{equation}
-{\cal L}_{EH}\simeq \frac{e^2\,f(v_F)}{96\pi \Delta} \, {\cal F}+\frac{e^4 \, g(v_F)}{256\pi \Delta^5} \, {\cal F}^2 \, .
\end{equation}
in which $f(v_F)=1-6v_{F}^2$ and $g(v_{F})=v_F^4-v_F^2+3/20$ are functions of the Fermi velocity.  
As example, for the graphene, the Fermi velocity in natural units is $v_{F}=0.0033$, such that these functions 
can be approximated by $f\simeq 0.99$ and $g\simeq 0.15$. The effective lagrangian is

\begin{eqnarray}\label{LPEHext}
\mathcal{L}_{PEH} 
&=&
-\frac{1}{4} \, F_{\bar{\mu}\bar{\nu}} \frac{2}{\sqrt{\bar{\Box}}} F^{\bar{\mu}\bar{\nu}}
+\frac{\theta}{2} \, \epsilon^{\bar{\mu}\bar{\nu}\bar{\rho}} A_{\bar{\mu}}\frac{2}{\sqrt{\bar{\Box}}} \partial_{\bar{\nu}}A_{\bar{\rho}}
\nonumber \\
&&
\hspace{-0.5cm}
+\frac{e^2 \, f(v_F)}{192\pi \Delta} \,  G_{\bar{\mu}\bar{\nu}}G^{\bar{\mu}\bar{\nu}}
%\nonumber \\
%&&
%\hspace{-0.5cm}
-\frac{e^4 \, g(v_{F})}{1024 \pi \Delta^5} \, \left( G_{\bar{\mu}\bar{\nu}}G^{\bar{\mu}\bar{\nu}}\right)^2 \, ,
\hspace{0.5cm}
\end{eqnarray}
where $G^{\bar{\mu}\bar{\nu}}$ is the antisymmetric tensor defined by the matrix $G$ in (\ref{matrixG}), that satisfies the relation $G_{\bar{\mu}\bar{\nu}}G^{\bar{\mu}\bar{\nu}}=-2{\cal F}$. 
%We can define the $3$-potential $G^{\bar{\mu}}=(A^{0},v_{F}A^{i})$, 
%where the correspondent field strength tensor is $G_{\bar{\mu}\bar{\nu}}=\partial_{\bar{\mu}}G_{\bar{\nu}}-\partial_{\bar{\nu}}G_{\bar{\mu}}$. 
The quadratic corrections of $G_{\bar{\mu}\bar{\nu}}G^{\bar{\mu}\bar{\nu}}$ can be absorbed in the kinetic term if we redefine the potentials and fields as
\begin{subequations}
\begin{eqnarray}
A^{\bar{\mu}} \rightarrow \left[1-\frac{e^2\sqrt{\bar{\Box}}}{96\pi \Delta} \, f(v_F) \right]^{1/2} A^{\bar{\mu}} \; ,
\\
F^{\bar{\mu}\bar{\nu}} \rightarrow \left[1-\frac{e^2\sqrt{\bar{\Box}}}{96\pi \Delta} \, f(v_F) \right]^{1/2} F^{\bar{\mu}\bar{\nu}} \; , 
\end{eqnarray}
\end{subequations}
and substituting it in (\ref{LPEHext}), we neglect the terms of order $\theta\,e^2$ and $e^6$ to obtain the 
Euler-Heisenberg pseudo-ED lagrangian
\begin{eqnarray}\label{LPEHresult}
\mathcal{L}_{PEH} 
&=&
-\frac{1}{4} \, F_{\bar{\mu}\bar{\nu}} \frac{2}{\sqrt{\bar{\Box}}} F^{\bar{\mu}\bar{\nu}}
+\frac{\theta}{2} \, \epsilon^{\bar{\mu}\bar{\nu}\bar{\rho}} A_{\bar{\mu}}\frac{2}{\sqrt{\bar{\Box}}} \partial_{\bar{\nu}}A_{\bar{\rho}}
\nonumber \\
&&
\hspace{-0.2cm}
-\frac{e^4\,g(v_{F})}{1024 \pi \Delta^5} \left( G_{\bar{\mu}\bar{\nu}}G^{\bar{\mu}\bar{\nu}}\right)^2 \; .
\end{eqnarray}
From the lagrangian (\ref{LPEHresult}), the Schwinger's critical electric field is defined by
\begin{equation}\label{Ec}
E_{c} = \frac{\Delta^2}{e\,v_{F}}=174.17 \, \mbox{eV}^2= 2.68 \times 10^{6} \, \mbox{V}/\mbox{cm} \; ,
\end{equation}
and the critical magnetic field is
\begin{eqnarray}\label{Bc}
B_{c}=\frac{E_{c}}{v_F}=5.2 \times 10^{4} \, \mbox{eV}^2= 3.6 \times 10^{7} \, \mbox{T} \; ,
\end{eqnarray}
%
%
%\begin{eqnarray}
%E_{c} &=& \frac{2\sqrt{2}\,m^2}{\sqrt{\pi}\alpha \, v_F}\left(v_F^4-v_F^2+\frac{3}{20}\right)^{-1/2}=42337\mbox{MeV}^2
%\nonumber \\
%&&
%=1.86 \times 10^{20} \, \mbox{V}/\mbox{cm} \; ,
%\end{eqnarray}
%
where $v_{F}=0.003$, $e=\sqrt{4\pi \alpha v_{F}}=0.017$ and $\Delta=0.1$ eV for the example of the graphene, in which the effects of the EM field are sensible up to this scale.
The electric critical field is in same order of the typical values of the electric fields in Dirac materials \cite{Keser}. For the regime of validity of the 
(\ref{LPEHresult}), the electric and magnetic fields must satisfy the conditions $|{\bf E}| \ll E_{c}$ and $|{\bf B}| \ll B_{c}$, respectively.  

%The PEHED (\ref{LPEHresult}) remains gauge invariant, but the presence of the Fermi velocity parameter in the non-linear term breaks the Lorentz symmetry. 
%
%
%
%\begin{eqnarray}\label{LPEH}
%\mathcal{L}_{PEH} 
%&=&
%\frac{1}{2}\left( {\bf E} \cdot \frac{2}{\sqrt{\bar{\Box}}} {\bf E}-B\frac{2}{\sqrt{\bar{\Box}}}B\right)
%\nonumber \\
%&&
%\hspace{-0.5cm}
%+\frac{\theta}{2} V \frac{2}{\sqrt{\bar{\Box}}} B
%+\frac{\theta}{2} \left( {\bf A} \times \frac{2}{\sqrt{\bar{\Box}}}{\bf E} \right)_{z}
%\nonumber \\
%&&
%\hspace{-0.5cm}
%-\frac{\alpha \, v_{F}}{12m}\left(1-6v_{F}^2\right) \left({\bf E}^2-v_{F}^2 B^2\right)
%\nonumber \\
%&&
%\hspace{-0.5cm}
%-\frac{\pi\alpha^2\,v_F^2}{8m^5} \left( v_F^4-v_F^2+\frac{3}{20} \right) \left({\bf E}^2-v_{F}^2 B^2\right)^2 .
%\end{eqnarray}
%
Therefore, non-linear effects ruled by the quartic term in the tensor $G^{\bar{\mu}\bar{\nu}}$ of the effective lagrangian 
shows the emergence of a non-linear electrodynamics in $1+2$ dimensions. The EHPED (\ref{LPEHresult}) remains gauge invariant, 
but the presence of the Fermi velocity in the non-linear term breaks the Lorentz symmetry. We will investigate some of the effects 
caused by the non-linearity in the next section.

%
%***********************

%
%\begin{eqnarray}
%{\cal L}_{PEH} &=& -\frac{1}{4}\,F_{\bar{\mu}\bar{\nu}} \, \frac{2}{\sqrt{\bar{\Box}}} \,F^{\bar{\mu}\bar{\nu}}
%+\frac{\theta}{2} \, \epsilon^{\bar\mu\bar\nu\bar\rho} \, A_{\bar{\mu}} \, \frac{2}{\sqrt{\bar{\Box}}} \, \partial_{\bar{\nu} } A_{\bar\rho}
%\nonumber \\
%&&
%-\frac{e^2}{32\pi^2} \, \ln(m) \, ({\bf E}^2-\beta^2\,B^2)
%\nonumber \\
%&&
%-\frac{23\,e^4\,\beta^2}{768\pi^2 \, m^4} \, ({\bf E}^2-\beta^2\,B^2)^2 \; .
%\end{eqnarray}

%
%dimensionless Chern-Simons parameter, and 
%$j^{\bar{\mu}}$ is a classical source. As mentioned previously, the bar index runs as $\bar{\mu}=(0,1,2)$, the derivative operator is %$\partial_{\bar{\mu}}=(\partial_{t},\partial_{x},\partial_{y})$, and the bar D'Alembertian operator means %$\bar{\Box}=\partial_{\bar{\mu}}\partial^{\bar{\mu}}=\partial_{t}^2-\partial_{x}^2-\partial_{y}^2$ in $1+2$ dimensions. The limit $M \rightarrow \infty$ 
%reduces the $N$-operator in $N(\bar{\Box})=2/\sqrt{\bar{\Box}}$, and (\ref{LLWCS}) becomes the lagrangian of the pseudo-electrodynamics with a Chern-Simons term 
%discussed in the ref. \cite{Alves2}.  
%

%
\section{Dispersion relations of the PEHED in a external EM field}
\label{sec3}
Using the variational principle in the action of the effective lagrangian (\ref{LPEHresult}), we obtain the free field equations of the EHPED
\begin{eqnarray}
\nabla\cdot{\bf D}+\frac{\theta \, 2}{\sqrt{\bar{\Box}}}B=0 
\; , \; 
\epsilon_{ji}\partial^{i}H+\frac{ \theta \, 2}{\sqrt{\bar{\Box}}}\epsilon_{ji}E_{i}=\partial_{t}D_{j} 
\; , \; \; \;
\end{eqnarray}
where ${\bf D}$ and $H$ are defined by
\begin{subequations}
\begin{eqnarray}
{\bf D} &=& \frac{2}{\sqrt{\bar{\Box}}}{\bf E}
%-\frac{e^2}{24\pi m}(1-6v_F^2){\bf E}
%\nonumber \\
%&&
%\hspace{-0.6cm}
-\frac{e^4\,g(v_F)}{64\pi \Delta^5}
%\left(v_F^4-v_F^2+\frac{3}{20}\right)
\left({\bf E}^2-v_{F}^2B^2\right) {\bf E}  \; ,
\\
H &=& \frac{2}{\sqrt{\bar{\Box}}}B
%-\frac{e^2 \, v_F}{24\pi m}(1-6v_F^2)B
%\nonumber \\
%&&
%\hspace{-0.6cm}
-\frac{e^4 \, g(v_{F})}{64\pi \Delta^5} \left({\bf E}^2-v_{F}^2B^2\right) v_{F} B \, .
\end{eqnarray}
\end{subequations}
These quantities show explicitly that we have non-linear equations of ${\bf E}$ and ${\bf B}$ induced by the quantum corrections of the pseudo-QED. 
The Faraday-Lenz law remains the same one, {\it i.e.}, $\epsilon_{ij}\partial_{i}E_{j}=-\partial_{t}B$. 
We can exam optical properties of the EHPED in an EM background field. 
The prescription of the background field is so introduced writing the $3$-potential as
$A^{\bar{\mu}}(x)=a^{\bar{\mu}}(x)+A_{0}^{\,\bar{\mu}}(x)$, in which $a^{\bar{\mu}}(\bar{x})$ is the propagating potential over the plane, $A_{0}^{\,\bar{\mu}}(\bar{x})$ 
is the potential associated with the EM background. We consider the expansion up to second order in $a^{\bar{\mu}}$, where the quadratic 
fluctuations are important for the analysis of the propagation effects. The EM tensor $F_{\bar{\mu}\bar{\nu}}$ is decomposed as $F_{\bar{\mu}\bar{\nu}}(\bar{x})=f_{\bar{\mu}\bar{\nu}}(\bar{x})+F_{0\bar{\mu}\bar{\nu}}$, in which $f^{\bar{\mu}\bar{\nu}}=\partial^{\bar{\mu}}a^{\bar{\nu}}-\partial^{\bar{\nu}}a^{\bar{\mu}}=\left( \, -e^{i} \, , \, -\epsilon^{ij} \, b \, \right)$ is the strength field tensor for the EM propagating field, and $F_{0}^{\;\,\bar{\mu}\bar{\nu}}=\partial^{\bar{\mu}}A_{0}^{\;\,\bar{\nu}}-\partial^{\bar{\nu}}A_{0}^{\;\,\bar{\mu}} =\left( \, -E_{0}^{\,\,i} \, , \, -\epsilon^{ij} \, B_{0} \, \right)$ represents the background electric and magnetic fields.
If the background field is constant and uniform, we can write the relation $A_{0\bar{\mu}}(\bar{x})=-F_{0\bar{\mu}\bar{\nu}}\,x^{\bar{\nu}}/2$. The expansion of the Lagrangian (\ref{LPEHresult}) around the external field up to quadratic terms in 
the propagation fields is
\begin{eqnarray}
{\cal L}_{PEH} &\simeq& -\frac{1}{4} \, f_{\bar{\mu}\bar{\nu}} \frac{2}{\sqrt{\bar{\Box}}} f^{\bar{\mu}\bar{\nu}}
+\frac{\theta}{2} \, a_{\bar{\mu}}\frac{2}{\sqrt{\bar{\Box}}} \tilde{f}^{\bar{\mu}}
%\nonumber \\
%&&
%\hspace{-0.5cm}
%+\frac{\theta}{2} \, \tilde{F}_{0\bar{\mu}}\frac{2}{\sqrt{\bar{\Box}}}a^{\bar{\mu}}
%-\frac{e^2}{48\pi m}\left(1-6v_{F}^2\right)({\bf e}^2-v_F^2b^2)
\nonumber \\
&&
\hspace{-0.5cm}
-\frac{e^4 \, g(v_F)}{64 \pi \Delta^5} \left({\bf e}\cdot{\bf E}_{0}-v_{F}^2\,b\,B_{0}\right)^2 
%\nonumber \\
%&&
%\hspace{-0.5cm}
\!+{\cal O}({\bf e}^3,b^3) \; .
\hspace{0.5cm}
\end{eqnarray}
Therefore, the action principle yields the linearized equations 
\begin{eqnarray}\label{Eqsdh}
\nabla\cdot{\bf d}+\frac{\theta \, 2}{\sqrt{\bar{\Box}}}b=0 
\; , \; 
\epsilon_{ji}\partial^{i}h+\frac{ \theta \, 2}{\sqrt{\bar{\Box}}}\epsilon_{ji}e_{i}=\partial_{t}d_{j} 
\; , \; \; \;
\end{eqnarray}
where ${\bf d}$ and $h$ are given by 
\begin{subequations}
\begin{eqnarray}
{\bf d} &=& \frac{2}{\sqrt{\bar{\Box}}}{\bf e}
%-\frac{e^2}{24\pi m}(1-6v_F^2){\bf e}
%\nonumber \\
%&&
%\hspace{-0.6cm}
-\frac{e^4 \, g(v_F)}{32\pi \Delta^5} \, {\bf E}_{0}({\bf E}_{0}\cdot{\bf e})
\nonumber \\
&&
\hspace{-0.1cm}
+\frac{e^4 \, g(v_F)}{32\pi \Delta^5} \, v_{F}^2\,B_{0}\,{\bf E}_{0}\,b  \; ,
\\
h &=& \frac{2}{\sqrt{\bar{\Box}}}b
%-\frac{e^2 \, v_F}{24\pi m}\left(1-6v_F^2\right)b
%\nonumber \\
%&&
%\hspace{-0.6cm}
+\frac{e^4 \, g(v_F)}{32\pi \Delta^5}\, v_F^{4} \, B_{0}^2 \, b
\nonumber \\
&&
\hspace{-0.1cm}
-\frac{e^4 \, g(v_F)}{32\pi \Delta^5} \, v_F^{2} \, B_{0} \left({\bf E}_{0}\cdot{\bf e}\right) \, .
\end{eqnarray}
\end{subequations}
For free fields ${\bf e}$ and $b$, the plane wave solutions are
\begin{eqnarray}\label{waveplane}
{\bf e}({\bf r},t)={\bf e}_{0} \, e^{i ({\bf k}\cdot{\bf r}-\omega \, t)} 
\; , \;
b({\bf r},t)=b_{0} \, e^{i ({\bf k}\cdot{\bf r}-\omega \, t)} \; , 
\end{eqnarray}
with ${\bf r}=(x,y)$, ${\bf e}_0$ and $b_{0}$ are constants amplitudes, ${\bf k}$ is the wave vector, and $\omega$ is the frequency.
Substituting (\ref{waveplane}) in the eqs. (\ref{Eqsdh}), we obtain the relations among the amplitudes, ${\bf k}$-vector, $\omega$-frequency, 
and the background fields
%
%\begin{subequations}
\begin{eqnarray}\label{Eqsd0h0}
k_{i}\,d_{0i}=0
\; , \;
\epsilon_{ji}\,k^{i}\,h_{0}-i\,\theta\,\epsilon_{ji}\,e_{0i}=-i\omega\,d_{0j} \; ,
\end{eqnarray}
%\end{subequations}
%
where the amplitudes $d_{0j}$ and $h_0$ are defined by
\begin{subequations}
\begin{eqnarray}
d_{0i} &=& e_{0i}
%\nonumber \\
%&&
-\sqrt{-\bar{k}^2} \, \frac{e^4 \, g(v_F)}{32\pi \Delta^5} \, \left({\bf E}_{0}\cdot{\bf e}_{0}\right) E_{0i}
\nonumber \\
&&
+\sqrt{-\bar{k}^2} \, \frac{e^4\,g(v_F) }{64\pi \Delta^5} \, v_F^{2} \, B_{0} \, E_{0i} \, b_0 \; ,
\\
h_{0}&=& b_{0}-\sqrt{-\bar{k}^2} \, \frac{e^4 \, g(v_{F})}{64\pi \Delta^5}\, v_F^2 \, B_{0}\left({\bf E}_{0}\cdot{\bf e}_{0}\right)
\nonumber \\
&&
+\sqrt{-\bar{k}^2} \, \frac{e^4 \, g(v_{F})}{64\pi \Delta^5} \, v_{F}^4\, B_{0}^2\,b_{0} \; ,
\end{eqnarray}
\end{subequations}
and the relation $b_{0}=\epsilon_{ij}\,k_{i}\,e_{0j}/\omega$ is obtained from the Faraday-Lenz law.
The bar notation for the $k$-squared means $\bar{k}^2=\omega^2-{\bf k}^2$.
We can write these linearized amplitudes as : $d_{0i}=\varepsilon_{ij}\,e_{0j}$ and $h_{0}=\mu^{-1}b_{0}+\beta_{i}\,e_{0i}$, 
where the permittivity electric tensor and the permeability magnetic are, respectively, defined by
\begin{subequations}
\begin{eqnarray}
\varepsilon_{ij}(\omega,{\bf k}) &=& \delta_{ij}
-\sqrt{-\bar{k}^2} \, \frac{e^4 \, g(v_{F})}{64\pi \Delta^5} \, E_{0i} \, E_{0j} 
\nonumber \\
&&
\hspace{-0.5cm}
+\sqrt{-\bar{k}^2} \, \frac{e^4\,g(v_F)}{64\pi \Delta^5} \, v_F^{2} \, B_{0} \, E_{0i} \, \epsilon_{lj} \, \frac{k_{l}}{\omega}
\; ,
\\
\mu^{-1}(\omega,{\bf k}) &=& 1
%\nonumber \\
+\sqrt{-\bar{k}^2} \, \frac{e^4\,g(v_{F})}{64\pi \Delta^5} \, v_{F}^4 \, B_{0}^2 \; ,
\end{eqnarray}
\end{subequations}
with
\begin{eqnarray}
\beta_{i}=-\sqrt{-\bar{k}^2} \, \frac{e^4\,g(v_{F})}{64\pi \Delta^5} \, v_{F}^2 \, B_{0}\,E_{0i} \; .
\end{eqnarray}
Notice that, if $B_{0}=0$ or $E_{0i}=0$, we have $\beta_{i}=0$. As consequence of the relations (\ref{Eqsd0h0}), 
the electric wave amplitude is not perpendicular to the wave propagation direction. 
The eigenvalues of the permittivity matrix are given by 
\begin{subequations}
\begin{eqnarray}
\varepsilon_{1} &=& 1 \; , \;
\label{epsilon1}
\\ 
\varepsilon_{2} &=& 1-\frac{\sqrt{-\bar{k}^2}}{\omega}\frac{e^4g(v_{F})}{64\pi \Delta^5}\left[\,\omega\,{\bf E}_{0}^2+v_F^{2}\,B_{0}({\bf E}_{0}\times{\bf k})_{z} \,\right] \, ,
\hspace{1cm} 
\label{epsilon2}
\end{eqnarray}
\end{subequations}
and magnetic permeability is
\begin{eqnarray}\label{musol}
\mu(\omega,{\bf k}) = 1 -\sqrt{-\bar{k}^2} \, \frac{e^4\,g(v_{F})}{64\pi \Delta^5} \, v_{F}^4 \, B_{0}^2 \; ,
\end{eqnarray}
in which are real quantities in the frequency range of $\omega < |{\bf k}|$. The solution (\ref{epsilon1}) is the usual vacuum electric permittivity, while that the results (\ref{epsilon2}) and (\ref{musol}) show that the material medium is dispersive due to presence of the wave vector $(\bf k)$ and $\omega$-frequency. In fact, 
this dependency is exclusively due to non-locality of the EHPED, and also to the background fields. Turning off the background fields, the results reduce to 
the usual vacuum, {\it i.e.}, $\varepsilon_{1}=\varepsilon_{2}=\mu=1$. The non-linear effects of (\ref{epsilon2}) and (\ref{musol}) are relevant for EM background of
\begin{subequations}
\begin{eqnarray}
&&
|{\bf E}_{0}| \sim \sqrt{\frac{64\pi \Delta^5}{e^4\,g(v_{F})\,\omega}} \simeq 1.86 \times 10^{7} \, \mbox{V}/\mbox{cm} \, \left[ \frac{0.1\,\mbox{eV}}{\omega} \right]^{1/2} \!\! 
,
\label{E0value}
\nonumber \\
%\hspace{0.5cm}
%\\
%&&
%v_{F}^2 B_{0} \sim \sqrt{ \frac{64\pi m^5}{e^4\,g(v_{F}) \, \omega} }  
\\
&&
v_{F}^2 B_{0} \sim \sqrt{\frac{64\pi \Delta^5}{e^4\,g(v_{F})\,\omega}} \simeq  8.3 \times 10^{5} \, \mbox{T} \, \left[ \frac{0.1\,\mbox{eV}}{\omega} \right]^{1/2} 
\!\! , \hspace{0.9cm}
\label{B0value}
\end{eqnarray}
\end{subequations}
when $v_{F} \ll 1$, and we have used the conversion of $\mbox{MeV}^2$ to V/cm and Tesla (T), respectively. For values of EM background compatible with 
the critical fields (\ref{Ec}) and (\ref{Bc}), we impose the lower bound $\omega > 0.1$ eV on the wave frequency.    
Substituting the Faraday-Lenz in the Ampère-Maxwell law in (\ref{Eqsd0h0}), the wave equation $M_{ij}e_{0j}=0$ for a general wave polarization $e_{0j}$, 
in which the symmetric wave matrix $M_{ij}$ is 
\begin{eqnarray}
&&
M_{ij}=\left[ 1+ \sqrt{n^2-1}\,\frac{e^4 \, g(v_F)}{64\pi \Delta^5} \, v_F^4 \, \omega \, B_{0}^2 \right] (n_{i}n_{j}-n^2 \delta_{ij})
\nonumber \\
&&
+\delta_{ij}+\sqrt{n^2-1} \, \frac{e^4\,g(v_F)}{64\pi \Delta^5} \, v_F^2 \, \omega \, B_0 \left(E_{0i}\epsilon_{jl}+E_{0j}\epsilon_{il}\right)n_{l}
\nonumber \\
&&
-\frac{i\theta}{\omega} \, \epsilon_{ij}-\sqrt{n^2-1} \, \frac{e^4\,g(v_{F})}{64\pi \Delta^5}\, \omega \, E_{0i} \, E_{0j} \; ,
\end{eqnarray}
where we have defined the refractive index $n=\sqrt{n_i\,n_i}$, with $n_{i}=k_i/\omega$. The solution of $M_{ij}e_{0j}=0$ is non-trivial when the determinant of $M_{ij}$ is null. In the general case, the determinant of $M_{ij}$ is
\begin{eqnarray}\label{detMij}
&&
\det(M_{ij})=1-n^2-\frac{\theta^2}{\omega^2}-\frac{e^4\,g(v_F)}{64\pi \Delta^5} \, v_F^2 \, \sqrt{n^2-1}
\nonumber \\
&&
\times \left\{ \omega \left({\bf E}_0 \times {\bf n}\right)^2+\omega\,v_F^2\,B_{0}^2\,n^2-2i\theta B_{0} \left({\bf E}_{0} \times {\bf n}\right)_{z} 
\right.
\nonumber \\
&&
\left.
-2\omega B_{0} \left[ E_{0x} \, n_{y}+E_{0y} \, n_{x}+n_x\,n_y ({\bf E}_{0}\cdot{\bf n}) \right]
\right\} \; .
\end{eqnarray}
When the electric background field is turned off $({\bf E}_{0}={\bf 0})$, the null determinant leads to equation
\begin{equation}\label{EqnB0}
n^2+\frac{\theta^2}{\omega^2}-1+\frac{e^4\,g(v_F)}{64\pi \Delta^5} \, v_F^4 B_{0}^2 \, \omega \, n^2\, \sqrt{n^2-1} = 0 \; .
\end{equation}
If we make $B_{0}=0$, the solution of (\ref{EqnB0}) is the refractive index of the pure Chern-Simons-Maxwell pseudo-ED :
\begin{eqnarray}\label{nCSM}
n_{cs}(\omega)=\sqrt{1-\frac{\theta^2}{\omega^2}} \; .
\end{eqnarray}
When $B_{0} \neq 0$, if we consider $\theta=0$ (without the CS term), the solutions 
for the refractive index are given by 
\begin{subequations}
\begin{eqnarray}
n_{1B}(\omega)= \sqrt{ \frac{1-\sqrt{1-4\,\ell_{B}^2\,\omega^2}}{2\,\ell_{B}^2\,\omega^2}  } \; ,
\label{n1B}
\\
n_{2B}(\omega)= \sqrt{ \frac{1+\sqrt{1-4\,\ell_{B}^2\,\omega^2}}{2\,\ell_{B}^2\,\omega^2}  } \; ,
\label{n2B}
\end{eqnarray}
\end{subequations}
where we have defined the length scale $\ell_{B}$ as
\begin{eqnarray}\label{ellB}
\ell_{B}\equiv\frac{e^4\,g(v_F)}{64\pi \Delta^5} \, v_F^4 B_{0}^2 \; .
\end{eqnarray}
The solutions (\ref{n1B})-(\ref{n2B}) are reals if the wave frequency satisfies the condition $\omega < (2\ell_{B})^{-1}$.
For a very small $\ell_{B}$, the solutions are reduced to
\begin{subequations}
\begin{eqnarray}
n_{1B}(\omega) &\simeq& 1+\frac{\ell_{B}^2\,\omega^2}{2}\simeq 1 \; .
\\
n_{2B}(\omega) &\simeq& \frac{1}{\ell_{B}\,\omega} = \frac{64\pi \Delta^5}{e^4 \,g(v_F) \,v_F^4 \, B_{0}^2\,\omega} \; .
\end{eqnarray}
\end{subequations}
When $\omega>(2\ell_{B})^{-1}$, the solutions provide the real and imaginary parts 
\begin{equation}
n_{1B}(\omega)=n_{2B}(\omega)=\frac{\sqrt{2\ell_{B}\,\omega+1}}{2\ell_{B}\,\omega}+i\,\frac{\sqrt{2\ell_{B}\,\omega-1}}{2\ell_{B}\,\omega} \; ,
\end{equation}
that shows a wave absorption in this range frequency.
In the presence of the CS parameter, the solutions are read as
\begin{subequations}
\begin{eqnarray}
n_{1B}(\omega) &\simeq& \frac{1}{\ell_{B}\,\omega}+\left(-1+\frac{2\theta^2}{\omega^2}\right)\frac{\ell_{B}\,\omega}{2} \; ,
\label{n1Btheta}
\\
n_{2B}(\omega) &\simeq& \sqrt{1-\frac{\theta^2}{\omega^2}}\left(1-\frac{i\ell_{B}\theta}{2} \right) \; ,
\label{n2Btheta}
\\
n_{3B}(\omega) &=& n_{2B}^{\ast}(\omega) \simeq \sqrt{1-\frac{\theta^2}{\omega^2}}\left(1+\frac{i\ell_{B}\theta}{2} \right) \; ,  
\label{n3Btheta}
\end{eqnarray}
\end{subequations}
in which $\omega > \theta$. Notice that in the solutions (\ref{n2Btheta}) and (\ref{n3Btheta}) the CS $\theta$-parameter 
contributes directly for the wave absorption in the material, where there is no any constraint involving the $\omega$-
frequency and the length parameter $\ell_{B}$. These results are compatible with the complex refractive index in graphene 
\cite{Wang}.

%the solution for the refractive index, in the case of small quantum corrections, is read as
%
%\begin{eqnarray}\label{n1B}
%n_{1B} (\omega) \simeq \frac{64\pi m^5}{e^4 \,g(v_F) \,v_F^4 \, \omega\,B_{0}^2} \; ,
%\end{eqnarray}
%
%where the result (\ref{nCSM}) is the contribution at lowest order in $e$.

%
In the case of a electric background field $(B_{0}=0)$, the null determinant in (\ref{detMij}) yields the equation
\begin{equation}\label{EqnE0}
n^2+\frac{\theta^2}{\omega^2}-1+\frac{e^4\,g(v_F)}{64\pi \Delta^5} \, v_F^2 \, ({\bf E}_0 \times \hat{{\bf k}})^2 \, \omega \, n^2 \sqrt{n^2-1}=0 \; .
\end{equation}
If ${\bf E}_{0}$ is parallel to $\hat{{\bf k}}$-direction, the solution of (\ref{EqnE0}) is given by (\ref{nCSM}). When we remove the CS parameter $(\theta=0)$, 
the solution for the refractive index is similar to (\ref{n1B}) and (\ref{n2B})  :
\begin{subequations}
\begin{eqnarray}
n_{1E}(\omega)= \sqrt{ \frac{1-\sqrt{1-4\,\ell_{E}^2\,\omega^2}}{2\,\ell_{E}^2\,\omega^2}  } \; ,
\label{n1E}
\\
n_{2E}(\omega)= \sqrt{ \frac{1+\sqrt{1-4\,\ell_{E}^2\,\omega^2}}{2\,\ell_{E}^2\,\omega^2}  } \; ,
\label{n2E}
\end{eqnarray}
\end{subequations}
where the length parameter $\ell_{E}$ now depends on the angle that the electric background field does with 
the wave propagation direction:
\begin{eqnarray}\label{ellB}
\ell_{E}\equiv\frac{e^4\,g(v_F)}{64\pi \Delta^5} \, v_F^2 \, ({\bf E}_{0}\times\hat{{\bf k}})^{2} \; .
\end{eqnarray}

If ${\bf E}_{0}$ is parallel to $\hat{{\bf k}}$, thus when $\ell_{E}=0$, the solution (\ref{n1E}) reduces to $n_{1E}=1$ (usual vacuum solution). 
For a very small $\ell_{E}$, the solution (\ref{n2E}) is
\begin{eqnarray}\label{n1E}
n_{2E}(\omega) \simeq \frac{1}{\ell_{E}\,\omega} \simeq \frac{64\pi \Delta^5}{e^4 \,g(v_F) \,v_F^2\,({\bf E}_{0} \times \hat{{\bf k}})^2 \, \omega} \; .
\end{eqnarray}
When $\theta \neq 0$, the refractive index has the solutions 
\begin{subequations}
\begin{eqnarray}
n_{1E}(\omega) &\simeq& \frac{1}{\ell_{E}\,\omega}+\left(-1+\frac{2\theta^2}{\omega^2}\right)\frac{\ell_{E}\,\omega}{2} \; ,
\label{n1Etheta}
\\
n_{2E}(\omega) &\simeq& \sqrt{1-\frac{\theta^2}{\omega^2}}\left(1-\frac{i\ell_{E}\theta}{2} \right) \; ,
\label{n2Etheta}
\\
n_{3E}(\omega) &=& n_{2E}^{\ast}(\omega) \simeq \sqrt{1-\frac{\theta^2}{\omega^2}}\left(1+\frac{i\ell_{E}\theta}{2} \right) \; ,  
\label{n3Etheta}
\end{eqnarray}
\end{subequations}
considering $\omega > \theta$. Notice that, if ${\bf E}_{0}$ is parallel to $\hat{{\bf k}}$, the wave absorption is null in (\ref{n2Etheta})-(\ref{n3Etheta}), 
all these solutions reduce to (\ref{nCSM}). We observe that both results (\ref{n1B}) and (\ref{n1E}) are proportional to the inverse of the $\omega$-frequency.
The origin of the wave absorption in the solutions is exclusively due to emergence of the radiative corrections at one loop for the fermionic action 
and the presence of the topological CS term in the gauge action. In next section, we discuss the birefringence in PEHED under external background fields.

\section{Birefringence in PEHED}
\label{sec4}
In this section, we discuss the birefringence phenomenon in the case of the pseudo-EH ED.  Since the EM plane wave propagates over the plane 
${\cal XY}$, the most general wave polarization is $ {\bf e}_{0}=(e_{0x} \, , \, e_{0y})$, as well as, the wave vector is $ {\bf k}=(k_{x} \, , \, k_{y})$. 
The birefringence analysis starts with particular considerations on the direction of the external field (magnetic or electric) and the 
wave polarization on the ${\cal X}$-direction, or on the ${\cal Y}$-direction. Since the PEHED is an planar theory defined on a space of $1+2$ dimensions, 
the external magnetic field is on the ${\cal Z}$-direction and the wave polarization vectors are on the bidimensional plane. If we change the direction of the wave 
polarization vectors, these vectors are perpendiculars to the external magnetic field in any situation. Thereby, the birefringence associated with the magnetic background field does not appear in $1+2$ dimensions. The case with electric background field is interesting due to electric vector is defined on the bidimensional plane. When $B_{0}=0$, we consider the situation of the electric background field on the ${\cal Y}$-direction, with  ${\bf E}_{0}=E_{0}\,\hat{{\bf y}}$. If we set the wave polarization as ${\bf e}_{0}=(e_{0x} \, , \, 0)$, the wave equation is $M_{11}\,e_{0x}=0$, and it reproduces a perpendicular refractive index $n_{\perp}$ as solution. When we have the situation of a linear wave polarization on ${\cal Y}$-direction, with ${\bf e}_{0}=(0 \, , \, e_{0y})$, the wave equation is $M_{22}\,e_{0y}=0$, in which it yields the solution of the parallel refractive index $n_{\parallel}$. The birefringence is defined by the subtraction :
\begin{eqnarray}\label{Deltan}
\Delta n = |n_{\parallel}-n_{\perp}| \; .
\end{eqnarray}  
%
%In the case of a pure magnetic background field $({\bf E}_{0}={\bf 0})$, the birefringence  the wave matrix $M_{ij}$ shows that the birefringence is null in $1+2$ dimensions, %{\it i.e.}, $\Delta n_{B}=0$. 
%
%
%
The wave equation $M_{11}\,e_{0x}=0$ yields the perpendicular refractive index $n_{\perp}=1$. If the wave polarization is on ${\cal Y}$-direction, the wave equation $M_{22}\,e_{0y}=0$ has the solution 
\begin{equation}
n_{\parallel}=\sqrt{1+\frac{e^8g^2(v_{F}) \, \omega^2E_0^4}{4096\pi^2\Delta^{10}}}\simeq 1+\frac{e^8g^2(v_F) \, \omega^2E_0^4}{8192\pi^2\Delta^{10}} \; .
\end{equation}
Using the definition (\ref{Deltan}), the electric birefringence is
\begin{eqnarray}\label{DeltanEE4}
\Delta n_{E}=\frac{e^8g^2(v_F) }{8192\pi^2\Delta^{10}} \, \omega^2  E_0^4 \; .
\end{eqnarray}
The result shows that the birefringence is proportional to wave frequency to square, and also is proportional to the electric background to fourth power.
If we substitute the values of $v_{F}=0.003$, $e=\sqrt{4\pi \alpha v_{F}}=0.017$ and $\Delta=0.1$ eV, we obtain the result for any electric background :
\begin{eqnarray}
\frac{\Delta n_{E}}{\omega^2 E_{0}^4} \simeq 2.33 \times 10^{-11} \, \mbox{eV}^{-10} \; .
\end{eqnarray}
Using the values of $E_{0}=1.2 \times 10^{3} \, \mbox{eV}^2$ (from eq. (\ref{E0value})) and $\omega \sim 0.1 \, \mbox{eV}$, we obtain the result
\begin{eqnarray}
\Delta n_{E}\simeq 0.48 \; ,
\end{eqnarray}
that has a value near to the strong birefringence in liquid crystals of graphene oxide \cite{Kim,Xu}. 
In optics, the electric birefringence is known as the Kerr effect, in which the variation of the refractive index in a 
medium changes with the applied electric field \cite{Kerr}. The known result for the Kerr effect for a nonlinear medium is $\Delta n = \lambda_{0} \, K \, |E_{0}|^2$, where $\lambda_{0}$ 
is the wavelength, $K$ is the Kerr constant for the medium, and ${\bf E}_{0}$ is the external electric field. First, we can compare it with the result (\ref{DeltanEE4}). The Kerr effect has quadratic 
dependency on the external electric field, while that the birefringence in (\ref{DeltanEE4}) is quartic in the electric field due to non-locality of the EHPED that appears since the original lagrangian of the model (\ref{LPEHresult}).
Thereby, the variation of refractive index in (\ref{DeltanEE4}) is more strong around the peak of intensity where the induced birefringence allows the creation of Kerr cells through the planar materials. 

%
%coupling of fermions with the Lee-Wick-Chern-Simons gauge field is introduced substituting the derivative operator 
%by the covariant derivative operator in the lagrangian (\ref{LDLW}) : $\partial_{\bar{\mu}} \mapsto D_{\bar{\mu}}=\partial_{\bar{\mu}}+i\,e\,A_{\bar{\mu}}$.
%The addition of the gauge sector (\ref{LLWCS}) leads us to the LWCSPQED 
%

%
%where $\slashed{D}=\Gamma^{\bar{\mu}}D_{\bar{\mu}}$ is the covariant derivative operator contracted 
%with the $\Gamma^{\bar{\mu}}$-matrices, and the coupling constant $(e)$ is dimensionless in $1+2$ dimensions, 
%that is related to fine structure constant by $e^2=4\pi \, \beta \, \alpha$, in which $\alpha=1/137$ is the fine structure constant.
%This lagrangian is $U(1)$-local gauge invariant, but the fermion Lee-Wick term contains the covariant derivative 
%operator in third order.

%

%
\section{Conclusions}
\label{sec5}
In this paper, the effective action of the fermion sector in the pseudo-quantum electrodynamics (PQED) is calculated to generate the so called Euler-Heisenberg pseudo-electrodynamics (EHPED). It is a non-linear (and also a non-local) electrodynamics defined on the planar space of $1+2$ dimensions. The Chern-Simons (CS) topological term is added to achieve a complete PED gauge invariant model. The fermions sector has a Lorentz symmetry violation in the PQED, this Lorentz symmetry breaking is transferred to the effective non-linear lagrangian. As consequence, the EHPED breaks the Lorentz symmetry due to presence of the Fermi velocity, that depends on the nature of the material medium. We explore the effects of the EHPED when the model is submitted to an external electromagnetic background, where these fields are considered uniform and constant. The electric permittivity tensor and magnetic permeability are calculated in terms of the EM external field. These results show that the planar material is a dispersive medium in which one of the eigenvalues of the electric permittivity tensor is function of the wave vector $({\bf k})$, and of the frequency $(\omega)$. The magnetic permeability is a scalar function in $1+2$ dimensions, that also is function of ${\bf k}$ and $\omega$. Posteriorly, we obtain the solutions for the refractive index of this planar material medium in the presence of a magnetic background field, and also for the case of only an electric background field, when the wave has a general polarization. The results show three solutions for the refractive index : The first is 
inversely proportional to wave frequency times the squared magnetic background field; the second and third solutions provide complex solutions, where the imaginary parts
are interpreted as wave absorptions that depend on the CS parameter, and on the radiative corrections emerged from the fermionic effective action. The case of a pure electric background field $({\bf E}_{0})$ is similar to magnetic case, but in the first solution, the refractive index depends on the angle that ${\bf E}_{0}$ does with ${\bf k}$. When ${\bf E}_{0}$ is parallel to ${\bf k}$-direction, the solutions are reduced to the refractive index of the CS pseudo-electrodynamics. Also in the electric case, 
two solutions exhibit real and imaginary parts with the wave absorption that are non null if ${\bf E}_{0}$ is not parallel to ${\bf k}$-direction. These results for the refractive index are compatible with the complex refractive index in the graphene, see the ref. \cite{Wang}.
For end, the birefringence is studied with the help of the wave equation in the EM background field. Since the external magnetic field has only component perpendicular to the planar space, the magnetic birefringence does not appear in $1+2$ dimensions. The birefringence arises only when we consider the electric background field, in which we investigate it for linear wave polarization. The non-local structure of the EHPED leads to a birefringence proportional to $\omega^2 E_{0}^{4}$, whose effect is of the order of $\sim 2.33 \times 10^{-11} \, \mbox{eV}^{-10}$. If we use a typical electric field of $E_{0}=10^{3} \, \mbox{eV}^2$ in a frequency of $\omega\sim 0.1 \, \mbox{eV}$, the permittivity and permeability of the planar medium is relevant, and the electric birefringence is estimated by $\Delta n_{E}\simeq0.48$. The result for the electric birefringence in the EHPED is different in relation to the usual effect that is known as the Kerr effect. Basically, the non-locality of the 
PEHED induces a birefringence that depends on the external electric field elevated to fourth potency. The Kerr effect implies into variation of the refractive index proportional to the squared of electric field. Thus, the electric birefringence in the EHPED is strong around the peak of intensity $E_{0}^4$. As application of these results, planar materials can be used as Kerr cells to modulate or rapidly interrupt the passage of a light beam. In the context of the dimensional reduction that led to PED and of the optical properties, others non-linear electrodynamics known in the literature under the prescription of the EM external fields can be investigated in a forthcoming project.

\section*{Acknowledgments}
The author expresses his gratitude to José A. Helay\"el-Neto for valuable discussions and suggestions on the aspects of the Euler-Heisenberg pseudo-ED.
This work was supported by the Conselho Nacional de Desenvolvimento Cient\'ifico e Tecnol\'ogico (CNPq) under grant 305743/2026-0 (GM).

\end{document}